\documentclass[12pt]{article} 

\usepackage{hyperref}
\usepackage{url}
\usepackage{setspace}
\usepackage{amsmath,amssymb,amscd,mathrsfs}
\usepackage{amsfonts}
\usepackage{braket}
\usepackage{color}
\usepackage{graphicx}
\usepackage{comment,tikz}
\usepackage{subcaption}

\numberwithin{equation}{section}

\def\p{\partial}
\newcommand{\bea}{\begin{eqnarray}}
\newcommand{\eea}{\end{eqnarray}}
\newcommand{\be}{\begin{equation}}
\newcommand{\ee}{\end{equation}}
\newcommand{\ba}{\begin{align}}
\newcommand{\ea}{\end{align}}

\newcommand{\CC}{\mathbb{C}} % Complessi
\newcommand{\C}{\mathcal{C}}

\newcommand{\Tr}{{\rm {Tr}}}

\newcommand{\cO}{\mathcal{O}}

\newcommand\rref[1]{(\ref{#1})}

\newcommand{\ie}{{\it i.e.~}}

\newlength{\slength}
\newcommand*{\bb}{\makebox[\slength][c]{$\bullet$}}
\newcommand*{\wb}{\makebox[\slength][c]{$\circ$}}

\newcommand{\ren}{R\'enyi\ }

\begin{document}

\begin{titlepage}
\begin{center}

\hfill \\
\hfill \\
\vskip 0.75in

{\Large \bf  Genus Two Partition Functions and \ren Entropies of Large $c$ CFTs}\\

\vskip 0.4in

{\large Alexandre Belin${}^a$, Christoph A.~Keller${}^b$, and Ida G.~Zadeh${}^b$}\\
\vskip 0.3in

${}^{a}${\it Institute for Theoretical Physics, University of Amsterdam,
Science Park 904, Postbus 94485, 1090 GL Amsterdam, The Netherlands} \vskip .5mm
${}^{b}${\it Department of Mathematics, ETH Zurich, CH-8092 Zurich, Switzerland} \vskip .5mm

\texttt{a.m.f.belin@uva.nl, christoph.keller@math.ethz.ch, zadeh@math.ethz.ch}

\end{center}

\vskip 0.35in

\begin{center} {\bf ABSTRACT } \end{center}
We compute genus two partition functions in two dimensional conformal field theories at large central charge, focusing on surfaces that give the third \ren entropy of two intervals. We compute this for generalized free theories and for symmetric orbifolds, and compare it to the result in pure gravity. We find a new phase transition if the theory contains a light operator of dimension $\Delta \leq 0.19$. This means in particular that unlike the second \ren entropy, the third one is no longer universal. 
\vfill

\noindent \today

\end{titlepage}

\tableofcontents

\section{Introduction}\label{sec_intro}
\subsection{Entanglement and the Ryu-Takayanagi formula}\label{subsec_intro_i}

Over the last decade, it has become clear that entanglement plays a key role in holography. At the beginning of this development was the proposal of Ryu and Takayanagi, that entanglement in the CFT is given by a minimal surface in the dual gravity theory \cite{Ryu:2006bv,Ryu:2006ef,Nishioka:2009un}. Proven a few years later \cite{Lewkowycz:2013nqa}, this proposal initiated a holographic entanglement program that has underlined the importance of entanglement in building a classical spacetime \cite{VanRaamsdonk:2009ar, VanRaamsdonk:2010pw,Bianchi:2012ev}. Results include the derivation of Einstein's equation from entanglement \cite{Lashkari:2013koa,Faulkner:2013ica}, reconstruction of bulk regions known as entanglement wedges \cite{Jafferis:2015del, Dong:2016eik, Faulkner:2017vdd, Cotler:2017erl} and much more. Entanglement has also played an important role just within field theory, as for example shows the recent derivation of the averaged null energy condition \cite{Faulkner:2016mzt} and the a-theorem \cite{Casini:2017vbe} using entanglement.

Part of the power of the Ryu-Takayanagi (RT) formula lies in the fact that entanglement is notoriously difficult to compute in quantum field theory. This is yet another example of the beauty of holography, where a quantity that is hard to compute in the field theory can be much simpler in the bulk. Still, in this article we want to compute entanglement measures on the boundary, using the holographic dual CFT.

The reason for this is the following.
As usual we will consider the correspondence in the large $N$ limit, so that the theory of gravity is weakly coupled. This does not mean that the theory is automatically some kind of perturbative Einstein gravity. In particular, such a weakly coupled theory can have a moduli space, which for instance can be explored by changing the string length. The `gravity point' is the point where the string length goes to zero, so that stringy effects do not play a role. Gravity at that point is then simply perturbative Einstein gravity, possibly coupled to a few light fields. A priori the RT computation is only reliable near that gravity point. If one moves away from that point, for instance to a point where stringy states begin to play an important role, this type of computation is no longer necessarily trustworthy. As we are interested also in exploring entanglement of stringy theories, we are thus forced to rely on CFT methods.

More precisely, we will consider two points on the moduli space: first, the gravity point discussed in the previous paragraph. Here we should really study the CFTs when they are strongly coupled. But instead, we model them by taking them to be dominated by Virasoro descendants and possibly some light fields. We can directly compare these results to the dual gravity theory and check if all expectations match.

The second point on the moduli space which we study is the symmetric orbifold point. Take for instance the AdS$_3$/CFT$_2$ correspondence coming from the D1-D5 system on $K3$ or $\mathbb T^4$. Here we know that there is a point on the moduli space where the CFT is a symmetric product orbifold theory \cite{Vafa:1995bm,Dijkgraaf:1998gf,Seiberg:1999xz,Larsen:1999uk} (see \cite{David:2002wn} for a review). At that point we are in a highly stringy regime \cite{Gaberdiel:2014cha}, so there is no direct gravity computation against which we can check the CFT results. Still, we can investigate to what extent the results change across the moduli space, or if there are possibly non-renormalization theorems at play.

\subsection{Entanglement \ren entropies and phase transitions}\label{subsec_intro_ii}
The most frequent way of computing entanglement entropy in the CFT is via the replica trick \cite{Calabrese:2004eu,2009JPhA...42X4005C} where one starts by computing the entanglement \ren entropies given by path integrals on replicated manifolds. The idea is to compute the entanglement \ren entropies $S_{n}$ for all integer values of $n$ and then proceed to do an analytic continuation to obtain the entanglement entropy.
Note that the \ren entropies themselves also contain new information about the entanglement structure.

The simplest setup that is not completely fixed by conformal symmetry is the union of two disjoint intervals. Conformal symmetry implies that the result can only depend on the cross ratio of the four endpoints, which we denote by $y$. For a holographic theory, the RT formula then predicts that at the gravity point there is a phase transition at $y=1/2$, where two extremal surfaces in the bulk exchange dominance. Our goal is to investigate this phase transition from the CFT side.

This phase transition is well understood for the second \ren entropy. The observation is that in this case it can be mapped to the torus partition function of the CFT, as was done for instance in \cite{Headrick:2010zt}. It therefore only depends on the spectrum of the theory, which makes it easier to analyze. For instance it was shown that at the symmetric orbifold point the phase diagram is the same as at the gravity point, with a Hawking-Page transition at $y=1/2$ \cite{Keller:2011xi}. More generally, it is universal at large central charge provided the density of light states is sparse enough \cite{Hartman:2014oaa}.

This observation provides motivation to suggest a non-renormalization theorem on the moduli space for the phase diagram. Note, however, that even though the underlying theory may be supersymmetric, the quantities we compute are not. So, unlike for protected quantities such as the index of the BPS spectrum \cite{deBoer:1998us} and the structure constants of the chiral ring {\cite{deBoer:2008ss,Baggio:2012rr} --- for which there is actual agreement at different points on the moduli space \cite{Gaberdiel:2007vu,Pakman:2007hn,Taylor:2007hs} --- there is no reason from supersymmetry to expect the results to agree. Similarly, in higher dimensions the free energy is not invariant on the moduli space, as shown in various examples \cite{Aharony:2003sx,Aharony:2005ew,Belin:2016yll}. Still, the universality of the $d=2$ results for the $n=2$ \ren entropy may motivate the formulation of a non-renormalization theorem as suggested in \cite{Headrick:2010zt}.

Computing higher \ren entropies is much harder.
At the gravity point, they can be computed directly in the bulk \cite{Faulkner:2013yia}. This is done by assuming that the bulk geometries preserve replica symmetry and by evaluating the on-shell action. On the CFT side, the issue is that compared to the second \ren entropy, now also the OPE coefficients contribute. This makes the computation much harder, and can potentially change the story dramatically. Concretely, the \ren entropies can be calculated by studying correlation functions of twist operators in a $\mathbb Z_n$ orbifold CFT \cite{Headrick:2010zt}.  In \cite{Hartman:2013mia}, it was shown that if the identity block dominates this correlation function, then the answer is universal and only depends on $c$ to leading order. 
If one makes the additional assumption that the identity block dominates for all $y<1/2$, then \cite{Hartman:2013mia} showed that the answer reproduces the gravity calculation of \cite{Faulkner:2013yia}. This is of course a very strong assumption, and the goal of our paper is to investigate under which condition it holds.

\subsection{Summary of Results}\label{subsec_intro_iii}

In this paper, we compute the third \ren entropy of two disjoint intervals at both the gravity point and the symmetric orbifold point which we discussed in section \ref{subsec_intro_i}. Surprisingly, this will be tractable even in the highly stringy regime of symmetric orbifold theories. This is due to the fact that both regimes of the theory share a common feature: they obey large $N$ (here large $c$) factorization \cite{ElShowk:2011ag}, which was shown for symmetric product orbifolds in \cite{Belin:2015hwa}. As we show here, this means that to leading order in $c$, the theory looks like a theory of generalized free fields and the OPE coefficients can be obtained by counting Wick contractions. This will allow us to estimate the dominant contribution to the genus two partition function, and hence to the third \ren entropy.

To compute the third \ren entropy for two disconnected intervals as a function of the cross-ratio $y$, we evaluate the genus two partition function. We show that for small $y$ it is given by
	\be
	Z_2(y) \simeq \sum_{\phi_1,\phi_2,\phi_3}|C_{\phi_1\phi_2\phi_3}|^2 \left(\frac{y}{27}\right)^{\Delta_1+\Delta_2+\Delta_3} \,,
	\ee
where we sum over all operators $\phi_i$ in the theory, with $C_{\phi_1\phi_2\phi_3}$ being their three point functions, and $\Delta_i$ their conformal weight. As advertised above, unlike for genus one, the genus two partition function depends on the OPE coefficients. If a given CFT is to have a universal third \ren entropy which agrees with the expected gravity result, this expression must converge for all $y<1/2$ as $c\to \infty$. We will show that this is not universally true.
	
In order to do so, we will calculate the OPE coefficients in the large $c$ limit. We find that the dominant contribution to the partition function is given by
	\be
	C_{\Delta_1\Delta_2\Delta_3} \sim 2^{\frac{\Delta_1+\Delta_2+\Delta_3}{2\Delta_{\min}}}\ .
	\ee
In a generalized free theory such as the dual CFT at the gravity point, $\Delta_{\min}$ is the weight of the lightest --- non-identity --- field. The result for the symmetric orbifold theory turns out to be the same, with $\Delta_{\min}$ the weight of the lightest single trace operator in the theory. It follows then that if the theory has a light enough field, to be precise with
	\be
	\Delta_{\min} \leq \Delta_0 \approx 0.19 \,,
	\ee
the genus two partition function will diverge at some $y_c<1/2$, giving a new phase. Unlike the second \ren entropy, the third \ren entropy is thus no longer universal. This is the main result of the present paper. Note that this results relies solely on large $c$ factorization, which is present in symmetric product theories but also in CFTs dual to Einstein gravity with some scalar fields.
	
In particular, this predicts that the computation done by Faulkner  in pure gravity \cite{Faulkner:2013yia} can be drastically changed in the presence of a (light) scalar field, and predicts the existence of a new solution to the equations of motion. This is reminiscent of a result obtained in higher dimensions for the \ren entropies of a spherical region \cite{Belin:2013dva} where a light scalar operator can condense for sufficiently large \ren index $n$. In two dimensions, the \ren entropies of a single interval are fixed by conformal symmetry but our results indicate that for two intervals, there appears to be a similar phenomenon as in higher dimensions. This is particularly interesting as the CFT interpretation of the results in \cite{Belin:2013dva} is still unknown.

\section{Holographic entanglement and \ren entropy}\label{sec_renyi}
\subsection{Entanglement and \ren entropies in 2d CFTs}\label{subsec_renyi}

We are interested in entanglement and \ren entropies in holographic CFTs. We start by a review of the basic definitions relevant for this work. Consider a CFT $\mathcal{C}$ with Hilbert space $\mathcal{H}$ and a state $\ket{\psi}\in\mathcal{H}$. Imagine splitting the Hilbert space in two as
\be
\mathcal{H}=\mathcal{H}_A\otimes \mathcal{H}_B \,.
\ee
The reduced density matrix is defined as
\be
\rho_A=\Tr_{\mathcal{H}_B} \ket{\psi}\bra{\psi} \,.
\ee
It encodes the entanglement between the subsystems $A$ and $B$. A standard measure of entanglement is the entanglement entropy, given by taking the von Neumann entropy of the reduced density matrix:
\be
S_{EE}(A)= -\Tr \rho_A \log \rho_A \,.
\ee
Computing the entanglement entropy directly is usually difficult in a quantum field theory as it involves computing the logarithm of the reduced density matrix. Nevertheless, it can be computed by means of the replica trick \cite{Calabrese:2004eu,2009JPhA...42X4005C}. One starts by considering the entanglement \ren entropies, defined as
\be
S_n = \frac{1}{1-n} \log \Tr \rho_A^n \,,
\ee
and, assuming that the \ren entropies are analytic in $n$%\footnote{Note that there are known examples where non-analyticities appear in the context of holography \cite{Belin:2013dva,Belin:2014mva}.}
, one gets the entanglement entropy by taking the limit
\be
S(A)=\lim_{n\to1} S_n \,.
\ee
Note that computing the \ren entropies is interesting in its own right as they give the complete basis-independent description of the reduced density matrix (being the moments of its eigenvalue distribution).

In this paper, we will be interested in probing spatial entanglement, so we will consider splitting a constant time slice of the CFT into two parts. The simplest example is to consider one interval of length $R$. In two dimensional CFTs, the \ren entropies are fixed by conformal symmetry \cite{Calabrese:2004eu,2009JPhA...42X4005C} and read
\be
S_n=\frac{c}{6}\left(1+\frac{1}{n}\right) \log \frac{R}{\epsilon} \,,
\ee
where $\epsilon$ is the UV cut-off. The entanglement entropy is then found to be
\be
S_A=\frac{c}{3} \log \frac{R}{\epsilon} \,.
\ee
Note that these quantities are UV-divergent in a quantum field theory, because of short-distance correlation through the entangling surface. 

We see that this configuration does not give much information on the theory. We therefore want to consider the next simplest configuration, where we take the region $A$ to be the union of two disjoint intervals with end points $(z_1,z_2)$ and $(z_3,z_4)$, respectively. To compute the \ren entropy $S_n$, one must compute the path integral over a complicated manifold, in this case a genus $n-1$ surface. The replica trick allows one to compute this quantity in terms of a correlation function in the orbifold theory $\mathcal{C}^{\otimes n}/\mathbb{Z}_n$
\be\label{Renyi}
\Tr \rho_A^n \propto \braket{\sigma_n(z_1)\bar{\sigma}_n(z_2)\sigma_n(z_3)\bar{\sigma}_n(z_4)} \,,
\ee
where the $\sigma_n$ are twist-$n$ operators in the orbifold theory. Because of conformal symmetry, this correlation function only depends on the cross-ratio of the four points, so we can fix three of these points to lie at $0,1$ and $\infty$, with only one remaining free point at $y$ with $0<y<1$. The \ren entropy (\ref{Renyi}) is then given by the partition function of the original theory $\mathcal{C}$ on a genus $n-1$ surface. The moduli of this surface only depend on the one parameter, $y$, so that we are on a special locus of the moduli space of genus 2 Riemann surfaces. 

Sometimes it is more convenient to consider UV-finite quantities. One such quantity is the mutual information. If we call our two disjoint intervals $A_1$ and $A_2$. The mutual information is given by
\be
I(A_1,A_2)=S(A_1)+S(A_2)-S(A_1\cup A_2) \,,
\ee
and, similarly, the mutual \ren information is defined as
\be\label{MRI_i}
I^{(n)}(A_1,A_2)=S_n(A_1)+S_n(A_2)-S_n(A_1\cup A_2) \,.
\ee
Note that the relevant correlation function to compute $I^{(n)}(A_1,A_2)$ is
\be\label{MRI}
\frac{\braket{\sigma_n(z_1)\sigma_n(z_2)\sigma_n(z_3)\sigma_n(z_4)} }{\braket{\sigma_n(z_1)\sigma_n(z_2)}\braket{\sigma_n(z_3)\sigma_n(z_4)}} \,.
\ee
In what follows, the distinction between (\ref{Renyi}) and (\ref{MRI}) is not too important. Note that the two differ only by theory independent factors, as the two point functions in the denominator of (\ref{MRI}) only depend on the central charge of $\C$ and $n$. For our purposes their only effect is to normalize the vacuum contribution to one, and to cancel an overall contribution coming from a Weyl rescaling when going to the covering surface (see \cite{Headrick:2010zt} and appendix D of \cite{Headrick:2015gba} for more details).

Computing the four point function in the numerator of (\ref{MRI}), or alternatively, computing the partition functions on the Riemann surfaces is a very complicated task which has been performed explicitly only for relatively simple theories such as the free boson \cite{Calabrese:2009ez} and the Ising model \cite{Calabrese:2010he}. For $n=2$, the problem maps to a torus partition function which makes the situation much easier and can be done in some generality \cite{Headrick:2010zt}. In our case, \emph{i.e.}, for $n=3$, we will be able to compute it by using factorization properties in the large $N$ limit, which simplifies the computation significantly.

\subsection{Holography and the RT formula}\label{subsec_rt}
Although entanglement entropy is quite complicated to compute in field theory, Ryu and Takayanagi proposed that it can be computed easily for holographic CFTs, by considering the area of a minimal surface $\gamma_A$, extending into the bulk and homologous to the region $A$ \cite{Ryu:2006bv}:
\be
S_A= \frac{\text{Area of }\gamma_A}{4 G_N} \,, \qquad \partial\gamma_A=\p A \,.
\ee
The \ren entropies can either be computed by finding bulk solutions that have the genus $g$ surface as their boundary \cite{Faulkner:2013yia,Barrella:2013wja} or by considering the area of a cosmic brane sitting at the minimal surface \cite{Dong:2016fnf}. 

\begin{figure}[hbtp]
	\centering
	\begin{subfigure}[c]{0.45\linewidth}\hfil
		\raisebox{0.19\linewidth}{\includegraphics[width=\textwidth]{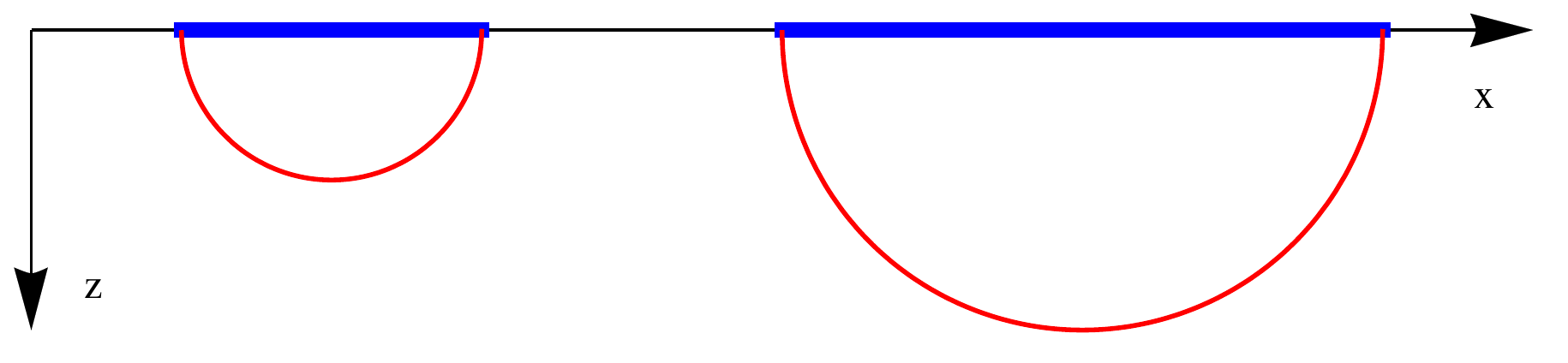}}
		\caption{The disconnected phase}
		\label{dim-lambda1}
	\end{subfigure}\hfill
	\begin{subfigure}[c]{0.45\linewidth}   
		\centering \includegraphics[width=\textwidth]{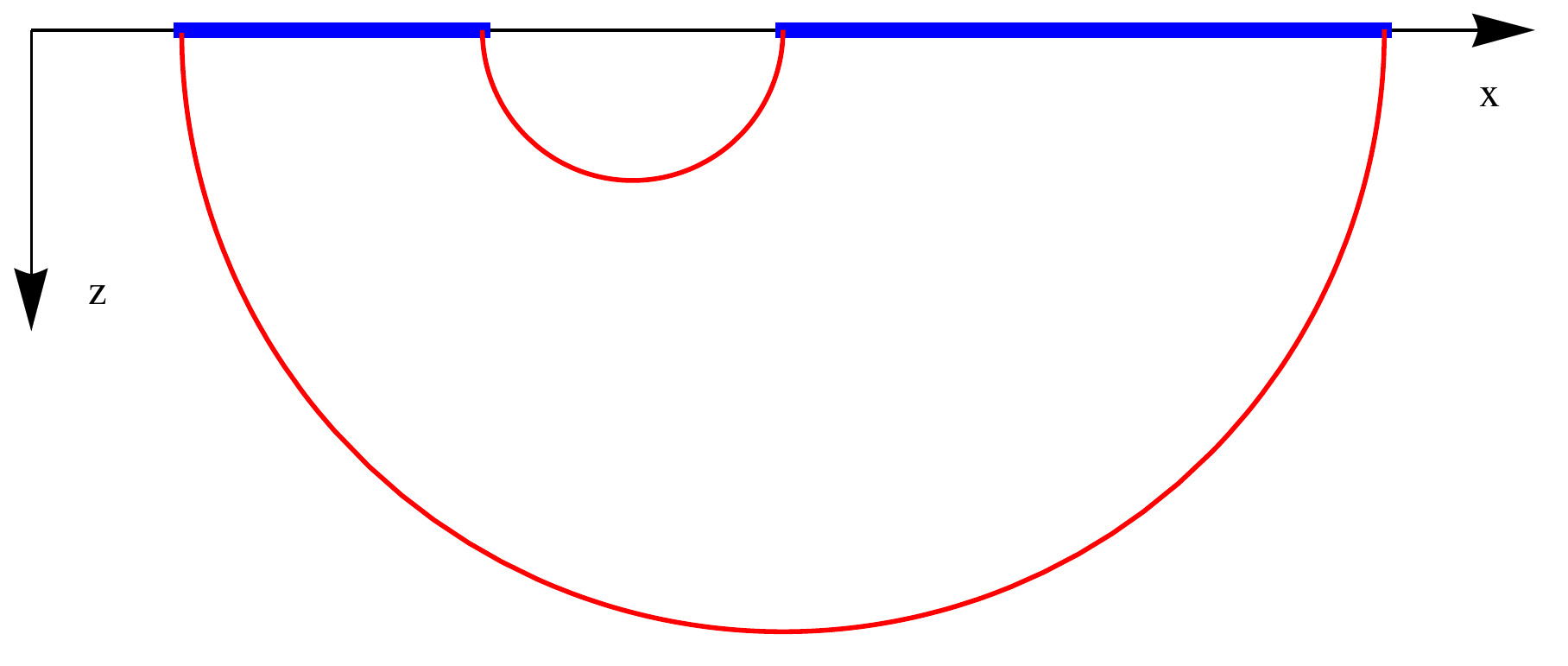}
		\caption{ The connected phase}
		\label{Tc_vs_Delta}
	\end{subfigure}
	\caption{The two different phases of entanglement entropy of two intervals, corresponding to different minimal surfaces in the bulk. The phase a) is dominant for $y<1/2$ and has $\mathcal{O}(1)$ mutual information. The phase b) dominates for $y>1/2$ and has $\mathcal{O}(c)$ mutual information}\label{2int}
\end{figure}

For two intervals, the entanglement entropy has a phase transition when $y=1/2$ as two different extremal surfaces exchange dominance (see figure \ref{2int}).  In \cite{Faulkner:2013yia}, assuming that the gravity solutions that preserve the replica symmetry were the dominant saddles, it was shown that the \ren entropies have a similar phase transition at $y=1/2$ where two saddles exchange dominance. There are of course many saddles that could contribute for $n>2$ but the results in \cite{Faulkner:2013yia} were obtained assuming none of the other solutions dominate (see also \cite{Hartman:2013mia}). Some numerical evidence for that claim was found in \cite{Maxfield:2016mwh}.

The CFT analysis performed in \cite{Headrick:2010zt,Hartman:2013mia} showed that the entanglement \ren entropies in holographic CFTs dual to pure gravity (\emph{i.e.}, theories which contain only the Virasoro vacuum block) exhibit a phase transition at $y=\textstyle\frac12$ for any value of $n\ge2$ and concluded that the same structure is preserved for the entanglement entropy under analytic continuation $n\to1$. 
The goal of this paper is to extend these results and compute the \ren entropies for $n=3$ in arbitrary large $c$ CFTs --- including the symmetric orbifold CFTs --- and compare with the gravity result.

\section{Genus one partition function}

\subsection{Phase transitions and radius of convergence}
As a warm-up, let us review the situation for the $n=2$ \ren entropy, that is the $g=1$ partition function
\be
Z(\tau)= \Tr_\mathcal{H} q^{L_0-\frac{c}{24}} \bar{q}^{\bar{L}_0 -\frac{c}{24}} = \Tr_\mathcal{H} e^{-\beta H + i \Omega J} \,.
\ee
Let us start out with the computation in pure gravity. For simplicity we will also set the angular potential $\Omega=0$. In that case  the contribution of the (classical) vacuum is 
\be
Z^{cl}_{vac}(\beta) = e^{\frac{c}{12}\beta}\ .\label{Z1vacpure}
\ee
Let us neglect any quantum corrections for the moment. It is clear that (\ref{Z1vacpure}) cannot be the correct answer, since it is not modular invariant under $\beta \mapsto 4\pi^2/\beta$. The missing piece is the contribution of the BTZ black holes, which gives 
\be
Z^{cl}(\beta) = e^{\frac{c}{12}\beta} + e^{\frac{c}{12}\frac{4\pi^2}{\beta}}\ .
\ee
To see the Hawking-Page transition in the large $c$ limit, we consider the rescaled free energy
\be
f(\beta) = -\frac{1}{c}\beta^{-1}\log Z\ ,
\ee
which is given by
\be\label{Z1pureIntro}
f(\beta) = \left\{\begin{array}{cc} - \frac{1}{12} &: \beta > 2\pi\\ - \frac{1}{12} \frac{4\pi^2}{\beta^2} &: \beta <2\pi \end{array} \right.\ .
\ee
We can indeed see the Hawking-Page transition at $\beta=2\pi$. 

On the other hand, when doing the computation on the CFT side, we do not rescale the free energy by $c$. Instead, to make sense of the large $c$ limit, we need to shift the vacuum energy to zero. For this reason, we define
\be
\tilde{Z}(\beta)= e^{- \beta \frac{c}{12}}Z(\beta) \,.
\ee
Similarly, we have
\be
\tilde{F}=-\beta^{-1} \log \tilde{Z} \,.
\ee
For finite (but very large) $c$ the answer is then
\be\label{F1pureIntro}
\tilde{F}(\beta) = \left\{\begin{array}{cc} 0 &: \beta > 2\pi\\  \frac{c}{12}(1-\frac{4\pi^2}{\beta^2}) &: \beta <2\pi \end{array} \right.
\ee
We see that the free energy diverges for $c\rightarrow\infty$ if $\beta<2\pi$. On the CFT side the phase transition is thus signaled by the fact that the partition function has a finite radius of convergence --- a Hagedorn transition on the CFT side thus signals the presence of a phase transition on the gravity side.

More generally, we see that the phase transition occurred because the only light state was the vacuum. If there are a large number of other light states, it is conceivable that their contribution could smooth out the phase transition at $\beta=2\pi$. From the CFT point of view this would be signaled by finding a divergence in the partition function for smaller temperatures already. A detailed analysis shows that this does not happen if the light spectrum is sparse enough \cite{Hartman:2014oaa}. The bound was found to be
\be \label{hksbound}
\rho(\Delta) \leq e^{2\pi \Delta} \,, \qquad \Delta\leq \frac{c}{12} \,.
\ee
In the remainder of this paper, we will only consider CFTs that satisfy this bound.

Including quantum corrections to pure gravity corresponds to including the contribution of the Virasoro descendants. The full contribution is then
\be \label{Z1pureQuant}
Z_{vac} = e^{\frac{c}{12}\beta}\prod_{n=2}^\infty \frac{1}{1-e^{-\beta n}}\ .
\ee
This growth due to descendants is much slower than \rref{hksbound}, so  it cannot affect the phase diagram and the free energy is still given by (\ref{Z1pureIntro}) and (to leading order) by (\ref{F1pureIntro}).

\subsection{The $N\to\infty$ limit}\label{ss:NLimit}

Let us now briefly discuss how we will compute our results. We will always work in the limit $N\rightarrow \infty$ from the start, that is we only keep the leading terms. For simplicity let us discuss the genus 1 case for the moment, where this corresponds to working with the density of states
\be
\rho_\infty(\Delta) := \lim_{N\to\infty}\rho_N(\Delta)
\ee
and considering the quantity $\tilde{Z}_\infty(\beta)= \sum_\Delta \rho_\infty(\Delta)e^{-\beta\Delta}$. $\rho_\infty(\Delta)$ is constructed by keeping the states with $\Delta$ fixed as $N\to\infty$. This can be thought of as keeping only the perturbative states of the dual gravitational theory.
We then investigate if $\tilde{Z}_\infty$ has a Hagedorn transition, that is we find its radius of convergence. If it diverges for some value of $\beta$, in view of (\ref{F1pureIntro}) we conclude that a phase transition has occurred, so that we are no longer in the vacuum phase.

Strictly speaking this raises an order of limit issue: we should really compute our quantities for large but finite $N$, and then take the $N\rightarrow\infty$ limit. 
Fundamentally the issue is that we should compute the quantity
\be
\lim_{N\to\infty} \tilde{Z}_N(\beta) = \lim_{N\to\infty} \sum_{\Delta}\rho_N(\Delta) e^{-\Delta\beta}\ .
\ee
What we are actually computing however is $\tilde{Z}_\infty(\beta)$, which only agrees with the quantity above if we can exchange the limit $N\to\infty$ with the infinite sum over $\Delta$.

Note that this is not a purely academic problem. Take for instance the example of pure gravity (\ref{Z1pureQuant}). In that case $\tilde{Z}_\infty$ is given by
\be
\tilde{Z}_\infty(\beta)=\prod_{n=2}^\infty \frac{1}{1-e^{-\beta n}}\ ,
\ee
since all the BTZ states are heavy and thus never show up after we have taken the $c\to\infty$ limit. The number of descendants only grows like $~e^{\sqrt{h}}$, which means that $\tilde Z_{\infty}$ converges for all $\beta$. We therefore do not find a Hagedorn transition, even though there clearly is a Hawking-Page transition at $\beta=2\pi$.
The situation is in fact similar for pure gravity at genus 2, as we do not see the expected Hagedorn transition at $y=1/2$.\footnote{Due to the growth of the three point functions, we do see a Hagedorn transition, but as it occurs at a higher value of $y$, it is not relevant and is merely an artifact of our order of limits.}

However, as long as we restrict to real values of the moduli this problem is usually not too severe. In that case all terms in the sum are positive, so that we can potentially apply various Lebesgue type theorems about exchanging sums and limits. Take for instance a case where $\rho_\infty(\Delta) \sim e^{\alpha \Delta}$. We can bound the partition function from below by restricting to a finite sum up to $\Delta_0$. For $\beta < \alpha$ we get
\be
\tilde{Z}_\infty(\beta) \geq \lim_{N\to\infty} \sum_{\Delta < \Delta_0} \rho_N(\Delta)e^{-\Delta\beta}
=\sum_{\Delta < \Delta_0} \rho_\infty(\Delta)e^{-\Delta\beta}
\sim e^{\Delta_0(\alpha-\beta)}\ ,
\ee
where we could exchange the limit because the sum is only finite. Since we can choose $\Delta_0$ arbitrarily big, this means that indeed $\tilde Z_\infty$ diverges for $\beta<\alpha$. Note that strictly speaking we would want a stronger result: Namely, that 
\be
\tilde{f}(\beta) = -\lim_{N\to\infty}\frac{1}{N}\log \tilde{Z}_N(\beta)
\ee
is non-zero. It could be that $\tilde{Z}_N(\beta)$, for instance, diverges like $N^{1/2}$, so that $\tilde{Z}_\infty$ diverges but $\tilde{f}(\beta)=0$. To avoid this issue it is necessary to know how quickly $\rho_N(\Delta)$ converges. For the symmetric orbifold for instance we know that $\rho_N(\Delta)=\rho_\infty(\Delta)$ if $\Delta< K N$ for some fixed constant $K$. It then follows that $\sum_{\Delta<KN} \rho_N(\Delta)e^{-\Delta\beta} \sim e^{KN(\alpha-\beta)}$, so that $f(\beta)$ is indeed non-zero.

Arguing the other way around is a bit more subtle, as indeed our counterexample shows. The basic issue is that in $\rho_\infty(\Delta)$ we are (by definition) neglecting all `heavy' states, that is states whose weight is of order $N$. It is possible however that those heavy states lead to a phase transition for all finite $N$. In order to avoid that, and to ensure that we are allowed to exchange the limit and the sum, we would like to apply some Lebesgue type theorem. In particular, if the density of states $\rho_N(\Delta)$ grows monotonically as a function of $N$ for all $\Delta$, we can apply the monotone convergence theorem. One such case is for instance the symmetric orbifold, where the number of states presumably increases monotonically. A case where this fails is pure gravity discussed above. If we fix the energy and vary $c$, the number of states depends on whether $\Delta$ is bigger or smaller than $c/12$. If it is larger, the number of states is given by the Cardy formula (BTZ states) and otherwise by the number of Virasoro descendants. In this case, one can see that the growth in $N$ is not monotonic. Nonetheless we will proceed under the assumption that the order of limits is not an issue, or at least that Hagedorn transitions of $\tilde Z_\infty$ tell us something interesting about phase transitions on the gravity side. 

In what follows we apply this approach to higher \ren entropies.
Even though these are of course well defined and finite for finite $N$, they may diverge as we take $N$ to infinity, just as in the genus 1 case discussed above. The radius of convergence of the $N=\infty$ theory then again, after an appropriate rescaling of the free energy by $N$, corresponds to a phase transition on the gravity side.

\section{Genus two partition function}\label{sec_g2}

\subsection{Higher genus partition functions}
Let us now turn to the genus two partition function which is relevant for the computation of the third mutual \ren information $I^{(3)}$ \rref{MRI_i}. Note that the torus partition function only depended on the spectrum of the theory. For higher genus, we will see that it depends on the other dynamical contents of the CFT: the OPE coefficients. 
As we will see, this changes the behavior quite a lot.

Schematically, we can express the partition function of a genus $g$ surface as
\be\label{genusg}
Z_g(\tau_i) \sim \sum_{\phi_i,\phi_j,\phi_k} \prod^{2g-2} C_{\phi_i\phi_j\phi_k} q_1^{h_1}\dots q_{3g-3}^{h_{3g-3}}\ .
\ee
Here we have decomposed the surface into $2g-2$ pairs of pants connected by $3g-3$ tubes, with $q_i$ the modulus of the $i$-th such tube. This is of course a very rough way of writing things, since the details of the coordinates we choose matter. In particular we have used the ordinary three point functions $C_{\phi\phi\phi}$ for the pair of pants contributions, whereas the exact contribution depends on which coordinates we choose. To relate this to the $(g+1)$-th Renyi entropy, we then also need the covering map to express the moduli $q_i$ in terms of the original cross ratio of the replica manifold, $y$.

Note that \rref{genusg} depends on two quantities:
\begin{enumerate}
	\item The spectrum of the theory, \ie the number of states $\rho(h)$ of a given weight $h$,
	\item The three point functions $C_{\phi_1\phi_2\phi_3}$.
\end{enumerate}

The reason why we are able to compute higher \ren entropies is that in the large $N$ limit, holographic theories should become free, as described for instance in \cite{ElShowk:2011ag}. As expected, this is exactly what happens for symmetric orbifolds \cite{Belin:2015hwa}. Due to this observation, computing higher genus partition functions in the large $N$ theory essentially reduces to a combinatorial problem of computing Wick contractions.

\subsection{Genus two partition function}\label{subsec_genus2}
We would like to compute the genus two partition function of 2d CFTs in the large central charge limit. 
For this we want to make (\ref{genusg}) more concrete. For simplicity we will in a first step only consider holomorphic theories.
We follow the approach of \cite{Gaberdiel:2010jf} and perform a Schottky uniformisation of the genus 2 Riemann surface. The partition function is then expressed in terms of four point functions on the sphere. We have
\be\label{partition_i}
Z_2=\sum_{h_1,h_2=0}^\infty C_{h_1,h_2}(x)\,p_1^{h_1}\,p_2^{h_2},
\ee
where $p_1$, $p_2$, and $x$ are the Schottky co-ordinates. The picture is that $x$ is the cross ratio of a sphere with 4 punctures, and $p_1$ and $p_2$ are the coordinates describing the two handles that glue the punctures together pairwise. We then sum over all fields running in the handles, which leads to the sum over $h_1$ and $h_2$. The partition function is defined on the Schottky space $\mathfrak{S}_2$,
\be
\mathfrak{S}_2 :=\{(p_1,p_2,x)\in\CC^3\mid x\neq 0,1,\
0<|p_i|<\min\{|x|,\,1/|x|\}\ , i=1,2\}\ .
\ee
The functions $C_{h_1,h_2}(x)$ are the four point functions on the sphere summed over all fields of weight $h_1$ and $h_2$:
\be\label{partition_iii}
C_{h_1,h_2}(x)=\sum_{\phi_i,\psi_i\in\mathcal H_{h_i}}K^{-1}_{\phi_1,\psi_1}K^{-1}_{\phi_2,\psi_2}
\Big\langle V^{out}(\bar\psi_1,\infty)\;V^{out}(\bar\psi_2,x)\;V^{in}(\phi_2,1)\;V^{in}(\phi_1,0)\Big\rangle\ .
\ee
Here $\bar\phi$ denotes the hermitian conjugate, and $K$ is the usual Kac matrix \cite{Dolan:1994st}.
The vertices are given by \cite{Gaberdiel:2010jf}
\bea\label{vinvout0inf}
&&V^{in}(\phi_1,0) = V(\phi_1,0) = \phi_1(0)\ ,\\
&&V^{out}(\psi_1,\infty)=
\lim_{z\to\infty}V(z^{2L_0}\,e^{z\,L_1}\psi_1,z)\ ,
\eea
and
\bea\label{vinvout1x}
&&V^{in}(\phi_2,1)=V\Big((x-1)^{L_0}e^{-L_1}\phi_2,1\Big)\ ,
\\
&&V^{out}(\psi_2,x)=V\Big((x-1)^{L_0}e^{L_1}\psi_2,x\Big)
\ .
\eea
These are the usual vertex operators when $\phi$ is a quasi-primary field. For non-quasi-primary fields however it is necessary to insert factors of $e^{L_1}$ in order to make them crossing symmetric. Note that the naive definition of the four point function is not crossing symmetric for non-quasi-primary fields, since they pick up additional transformation factors under the required M\"obius transformation.

By inserting a complete set of states $\phi_3$ and taking all states $\phi_1,\phi_2,\phi_3$ orthonormal, we can write (\ref{partition_i}) as 
\be\label{partition_v}
Z_2=\sum_{\phi_1,\phi_2,\phi_3} 
 C^{out}(\phi_1,\phi_2,\phi_3;x)C^{in}(\phi_1,\phi_2,\phi_3;x) p_1^{h_1} p_2^{h_2} x^{h_3-h_1-h_2}\ .
\ee
Here $C^{in}$ and $C^{out}$ are again closely related to the naive three point functions --- they are indeed obtained by inserting $V^{in}$ and $V^{out}$ at $z=1$ in the usual three point function. Note however that they do depend on $x$. They satisfy
\be
C^{out}(\phi_1,\phi_2,\phi_3;x)^*=C^{in}(\phi_1,\phi_2,\phi_3;x^*)\ ,
\ee
so that for $p_1$, $p_2$, and $x$ real, each term in (\ref{partition_v}) is manifestly non-negative. If $\phi_2$ is quasi-primary, then $C^{in}$ and $C^{out}$ reduce to the usual expression
\be
C^{in}(\phi_1,\phi_2,\phi_3;x) = (x-1)^{h_2} C_{\phi_1\phi_2\phi_3}\ .
\ee
Note that the invariance of the four point function under the M\"obius transformation $\hat\gamma=-1/z$ yields 
\be\label{xpower_ii}
C_{h_1,h_2}(x)=C_{h_1,h_2}\left(\frac1x\right)\ .
\ee
For future convenience we will thus work with $x^{-1}$ rather than $x$.

Let us now assume that all states are quasi-primary. This is of course not true, but for our computations we assume that this is a reasonable assumption. In particular, for the growth of states we will encounter, the number of non-quasi-primary fields will be subleading. It is quite possible that this approximation will change the precise values of the radius of convergences, but we believe that our results should be reasonably accurate. Alternatively, we can simply restrict the sum over $\phi_2$ to quasi-primary fields. By the remarks about positivity above, this will at least give a lower bound on the full result.
Working with $x^{-1}$ rather than $x$, equation (\ref{partition_v}) then leads to the simpler expression 
\be\label{Z2quasi}
Z_2\approx\sum_{\phi_1,\phi_2,\phi_3} |C_{\phi_1\phi_2\phi_3}|^2\, x^{-h_3}(1-x^{-1})^{2h_2}\left(p_1 x\right)^{h_1}\,\left(p_2 x\right)^{h_2}\ .
\ee

\subsection{The replica manifold}\label{subsec_genus2_replica}
We would like to express the partition function in terms of the co-ordinates on the genus 2 replica manifold. To do so, we derive the relation between the Schottky co-ordinates and the replica co-ordinates through matching the multiplicative periods of the two surfaces \cite{Headrick:2015gba}. We first consider the genus 2 Riemann surface uniformised by the Schottky group. The multiplicative periods are
\be\label{mltprd}
q_{ij}=e^{2\pi i\tau_{ij}},
\ee
where $\tau$ is the period matrix of the Riemann surface. The power series expansions of the multiplicative periods in terms of the Schottky co-ordinates are derived in equations (A.26)-(A.28) of \cite{Gaberdiel:2010jf}, and are of the form
\bea\label{mltprds}
&&q_{11}=p_1\sum_{n,m=0}^{\infty}\sum_{r=-n-m}^{n+m}c(n,m,|r|)\,p^n_1p^m_2x^r,\nonumber\\
&&q_{22}=p_2\sum_{n,m=0}^{\infty}\sum_{r=-n-m}^{n+m}c(m,n,|r|)\,p^n_1p^m_2x^r,\\
&&q_{12}=x+x\sum_{n,m=1}^{\infty}\sum_{r=-n-m}^{n+m}d(n,m,r)\,p^n_1p^m_2x^r,\nonumber
\eea
where the coefficients $c(n,m,|r|)$ and $d(n,m,r)$ are given in appendix E of the same reference. 

We next consider the replica manifold $\mathscr R_{2,n}$ where the superscript $2$ stands for 2 disjoint intervals and $n$ is the number of sheets. This manifold has genus $g=(n-1)$. The period matrix for $\mathscr R_{2,n}$ is derived in \cite{Calabrese:2009ez,Coser:2013qda} and is of the form
\be\label{rnprd}
\tau_{ij,\mathrm{rep}}=\frac{2i}{n}\sum_{k=1}^{n-1}\sin\left(\pi\frac{k}{n}\right)\,\cos\left(2\pi\frac{k}{n}(i-j)\right)\frac{_2F_1\left(\frac{k}n,1-\frac{k}n;1;1-y\right)}{_2F_1\left(\frac{k}n,1-\frac{k}n;1;y\right)},
\ee
where $y$ is the cross ratio of endpoints of the two intervals and the subscript ``rep" corresponds to the replica surface. The genus-2 surface $\mathscr R_{2,3}$ has just one modulus, $y$, and its period matrix is given by
\be\label{rnprd3}
\tau_{\mathrm{rep}}=\frac{2i}3\,\frac{_2F_1\left(\frac13,\frac23;1;1-y\right)}{_2F_1\left(\frac13,\frac23;1;y\right)}\left({\begin{array}{cc}
1 & -\frac12 \\
-\frac12 & 1 \\
\end{array} } \right).
\ee
The multiplicative periods for this surface are then given by $q_{\mathrm{rep}}=e^{2\pi i\tau_{\mathrm{rep}}}$.

In the short interval limit $y\ll1$, the multiplicative periods of the replica surface $\mathscr R_{2,3}$ are given by the power series expansions
\bea\label{rnmltprd}
&&q_{11,\mathrm{rep}}=e^{2\pi i\tau_{11},\mathrm{rep}}\big|_{y\ll1}=\frac{y^2}{729}+\frac{10y^3}{6561}+\frac{29y^4}{19683}+O(y^5),\nonumber\\
&&q_{12,\mathrm{rep}}=e^{2\pi i\tau_{12},\mathrm{rep}}\big|_{y\ll1}=\frac{27}{y}-15-2y-\frac{734y^2}{729}-\frac{4181y^3}{6561}+O(y^4),\\
&&q_{22,\mathrm{rep}}=q_{11,\mathrm{rep}},\qquad q_{12,\mathrm{rep}}=q_{21,\mathrm{rep}}.\nonumber
\eea

To find the relationship between the Schottky coordinates $\{p_1,p_2,x\}$ and the replica modulus $y$ we match their associated multiplicative periods (\ref{mltprds}) and (\ref{rnmltprd}), respectively. We observe that $q_{11,\mathrm{rep}}=q_{22,\mathrm{rep}}$, which yields $p_1=p_2\equiv p$. The short interval limit then corresponds to $p\ll1$. Reading the coefficients $c(n,m,|r|)$ and $d(n,m,r)$ from appendix E of \cite{Gaberdiel:2010jf} and inverting equations (\ref{mltprds}), we find \cite{Headrick:2015gba}:
\bea\label{pxy_sub}
p(y)\!\!\!&=&\!\!\!\frac{y^2}{729}+\frac{28}{19683}\,y^3+\frac{26}{19683}\,y^4+\frac{5768}{4782969}\,y^5 +\frac{47429}{43046721}\,y^6+\frac{10582844}{10460353203}\,y^7+\nonumber\\
&&\!\!\!\!\!\!\!\!\!\!+\;O(y^8),\nonumber\\
x(y)\!\!\!&=&\!\!\!\frac{27}{y}-15-\frac{56}{27}\,y-\frac{28}{27}\,y^2-\frac{12892}{19683}\,y^3-\frac{3044}{6561}\,y^4 +O(y^5)\,.
\eea
Note in particular that 
\be
p(y) = x(y)^{-2} + O(y^3)\ .
\ee
In the short interval limit of $y \ll 1$, we can plug (\ref{pxy_sub}) into (\ref{Z2quasi}) to get
\bea\label{Z2total}
&&Z_2\,\sim\sum_{\phi_1,\phi_2,\phi_3} |C_{\phi_1\phi_2\phi_3}|^2\, \left(\frac{y^2}{729}\right)^{h_1+h_2}\left(\frac{27}{y}\right)^{h_1+h_2-h_3}\\
&&\,\quad\;=\sum_{\phi_1,\phi_2,\phi_3} |C_{\phi_1\phi_2\phi_3}|^2\, \left(\frac{y}{27}\right)^{h_1+h_2+h_3}.\nonumber
\eea

We see that it is useful to define the coefficients
\be\label{define_D}
D(h)= \sum_{h_1+h_2+h_3=h} |C_{\phi_1\phi_2\phi_3}|^2\ ,
\ee
so that we can write the genus 2 partition function as the sum
\be
Z_2(y) \sim\sum_h D(h)\left(\frac{y}{27}\right)^{h}\ .
\ee
The radius of convergence of this expression is fixed by the growth behavior of $D(h)$. As we saw, those in turn depend on two factors: the growth of the three point functions $C_{\phi_1\phi_2\phi_3}$, and the number of corresponding states.

\subsection{Convergence at $y=1/2$}
So far, we have only discussed the case where $y$ is very small. This is of course enough to argue that there is a finite radius of convergence, \emph{i.e.}, that there is a phase transition in the first place. 
If we extrapolate from the Ryu-Takayanagi entanglement entropy, then the expected phase transition for pure gravity occurs at the self-dual point of crossing symmetry, that is at $y=1/2$. Let us check what this means for what we have said so far. The question is whether the partition function converges for $y=1/2$, which implies that we are still in the vacuum phase. If it diverges, however, it implies that there was a phase transition before that, \emph{i.e.}, that there is a new phase in the theory.

We do not have exact expressions for $p(\textstyle\frac12)$ and $x(\textstyle\frac12)$. Instead, we will use the expansions (\ref{pxy_sub}) up to the order indicated, which converge reasonably fast. This yields
\be
p\left(\textstyle\frac12\right) \approx e^{-7.31}\ ,\qquad x\left(\textstyle\frac12\right) \approx e^{3.63}\ . 
\ee
This shows that $p$ and $1/x$ are indeed very small, so that it is reasonable to trust our approximations.\footnote{In fact, one can show that at $y=1/2$ we have $x=-1+1/\sqrt{p}+\sqrt{p}$, which is compatible with our values. We thank Henry Maxfield for a discussion on this point.}
From this we get the expansion
\be\label{Z2sumdiffexp}
Z_2\left(\textstyle\frac12\right) \approx \sum_{h_1,h_2,h_3}|C_{\phi_1\phi_2\phi_3}|^2 e^{-3.68h_1 -3.73h_2-3.63h_3}\ .
\ee
Note that now the summand is no longer symmetric under exchanging the $h_i$, even though the total sum of course still is. It seems plausible that this asymmetry is due to various approximations we have made such as neglecting the effect of non-quasi-primaries. In any case if we assume for simplicity that all exponents are the same and just take their average, we then get the simpler expression
\be
Z_2\left(\textstyle\frac12\right) \approx \sum_h D(h)e^{-3.68 h}\ ,
\ee
where $D(h)$ is defined in (\ref{define_D}). This expression converges if $D(h)$ does not grow faster than $e^{3.68h}$. Note that this is relatively close to $e^{\pi h}$, which is of course very similar to the case of $g=1$, where the critical growth of the spectrum was $e^{2\pi h}$. A more detailed analysis of the maximal term contributing to $D(h)$ does not change this much, as can be seen in Appendix \ref{app:maxsum}.

\subsection{Non-holomorphic theories}
We shall now write the expression for the genus 2 partition function of non-holomorphic theories.  The computation is very similar to what we discussed above for holomorphic CFTs. The total conformal weight of an operator is given by 
\be
\Delta_i = h_i +\bar h_i\ .
\ee
The Schottky coordinates $p_1,p_2,x$ are still the same, but the partition function is no longer a meromorphic function of them. In particular, $Z_2(y)$ is then a non-meromorphic function. For simplicity, we will restrict to case where $y$ is real, since we are mostly interested in the case $y=1/2$ anyway. Also, as discussed in section \ref{subsec_genus2} (see the discussion below equation (\ref{xpower_ii})), we consider the contribution from quasi-primary states $\phi_1,\phi_2,\phi_3$. Equation (\ref{Z2total}) then becomes
\be\label{Z2_nonholom}
Z_2(y) \approx \sum_{\Delta_1,\Delta_2,\Delta_3}|C_{\phi_1\phi_2\phi_3}|^2 \left(\frac{y}{27}\right)^{\Delta_1+\Delta_2+\Delta_3}.
\ee
Note that for fixed $\Delta_i$ we sum over all fields of that weight, including over all spins. We then define again
\be\label{Delta}
D(\Delta)\equiv\sum_{\Delta_1+\Delta_2+\Delta_3=\Delta} |C_{\phi_1\phi_2\phi_3}|^2,
\ee
so that 
\be\label{Zathalf}
Z_2\left(\textstyle\frac12\right) \approx \sum_\Delta D(\Delta)e^{-3.68 \Delta}\ .
\ee

\section{The gravity point}\label{sec_grav}
In this section we compute the coefficient $D(\Delta)$ defined in (\ref{Delta}) for various theories that are of interest for holography. We first evaluate the OPE coefficients $C_{\phi_1\phi_2\phi_3}$ in subsection \ref{subsec_ff_C}, and will then compute the growth of $D(\Delta)$ in subsection \ref{sec_DDelta}.

\subsection{Growth of OPE coefficients}\label{subsec_ff_C}
We start by considering a CFT that describes weakly coupled gravity at the gravity point. This theory is very similar to pure gravity as it contains Virasoro descendants, as well as possibly a few additional light fields which become free in the $c\to\infty$ limit. This is very much a bottom up approach, as we do not know of a consistent explicit example of such a theory. However, since we are only interested in the light spectrum, this is not an issue here.

We consider therefore the correlation functions of a (generalized) free theory coming from a single free scalar $\phi$ of weight $\Delta_\phi$. For simplicity, we consider only operators with no derivatives\footnote{We discuss the case involving derivatives in appendix~\ref{app:derivatives}. Our results indicate however that the leading order behavior does not change.
}
, which are of the form
\be
\mathcal{O}=\mathcal{N}_\mathcal{O} : \phi^K :\ ,
\ee
where $\mathcal N_\mathcal{O}$ is the normalization constant. The two-point function is then given by all possible Wick contractions of $\cO$ with itself, of which there are $K!$, so that we get
\be
\langle :\phi^K:(z)\;:\phi^K:(0)\rangle = \frac{K!}{|z|^{2K\Delta_\phi}}\ .
\ee
The normalization constant is then found to be
\be\label{ff_normal}
\mathcal{N}_\mathcal{O} =\frac1{\sqrt{K!}}.
\ee

Let us now compute the three point function 
\be
C_{K_1K_2K_3}\equiv\big\langle \cO_1(\infty)\; \cO_2(1)\;\cO_3(0)\big\rangle
\ee
by again working out the combinatorics of the Wick contractions. Let us contract $J$ factors of $\phi$ in $\cO_1$ with $J$ factors of $\phi$ in $\cO_2$. Note that $J$ is actually determined by the condition that we need to contract all $K_i$ factors in all three operators: due to the constraint $K_1-J+K_2-J=K_3$ we then have
\be
J= \frac{1}{2}\,(K_1+K_2-K_3)\ .
\ee
The combinatorics of the three-point function is thus a question of how many different ways there are to distribute the Wick contractions. In $\cO_1$ we can pick the J factors that contract with $\cO_2$ in $\binom{K_1}{J}$ different ways, and in $\cO_2$ there are $\binom{K_2}{J}$ choices. We can then contract them with each other in $J!$ ways. The remaining factors in $\cO_1$ and $\cO_2$ are contracted with $\cO_3$, for which there are $K_3!$ different possibilities. Altogether we are thus left with 
\be
\frac{K_1! K_2! K_3!}{(K_1-J)!(K_2-J)!J!}
\ee
possible contractions. Taking into account the normalization factors $\mathcal{N}_{\cO_i}$ in (\ref{ff_normal}), we obtain
\be\label{C3ptfree}
C_{K_1K_2K_3}=\frac{\sqrt{K_1! K_2!K_3!}}{\left(\frac{K_1+K_2-K_3}{2}\right)!\left(\frac{K_1-K_2+K_3}{2}\right)!\left(\frac{-K_1+K_2+K_3}{2}\right)!}\ ,
\ee
which is symmetric in $K_1$, $K_2$, and $K_3$.

Let us now calculate the OPE coefficients of the CFT dual to pure gravity at large $c$ limit in the remainder of this section. To leading order in $c$, the result is again given by (\ref{C3ptfree}). The reason for this is that the OPE of $T$ with itself is
\be
T(z)T(0) \sim \frac{\frac c2}{z^4} + \frac{T(0)}{z^2}+ \frac{\partial T(0)}{z},
\ee
where $\sim$ corresponds to the singular terms. To leading order in $c$, this is the OPE of a free field with dimension $h=2$ and with normalization $\textstyle\frac c2$ rather than 1. When we properly normalize $T$, then to leading order in $c$ we automatically recover (\ref{C3ptfree}).

\subsection{Contribution to $D(\Delta)$}\label{sec_DDelta}

Let us now discuss the growth of $D(\Delta)$ in (\ref{Delta}). The conformal weight of the field $\mathcal O_i$ is given by $\Delta_i = \Delta_\phi K_i$. To obtain the maximal contribution to $D(\Delta)$, we thus want to maximize \rref{C3ptfree} under the constraint $\Delta = \Delta_\phi(K_1+K_2+K_3)$. To this end we use Stirling's formula
\be\label{stirling}
N! \sim \sqrt{2\pi N}\,N^N e^{-N}\ .
\ee
One can show that under the constraint of keeping $K=K_1+K_2+K_3$ fixed, equation \rref{C3ptfree} becomes maximal for $K_1=K_2=K_3$, giving 
\be
\left(\frac{\sqrt{K_1!}}{\frac{K_1}{2}!}\right)^3 \sim 2^{\frac{3K_1}2}\ ,
\ee
growing exponentially in $K_1$  --- see appendix~\ref{app:maxsum}. 
According to equation (\ref{Delta}), the contribution of this term to $D(\Delta)$ is then 
\be\label{Dfree}
D(\Delta) \sim 2^{\Delta/\Delta_\phi}\ .
\ee
Note that to the precision that we are interested in, the maximal term in the sum will give the leading behavior, and including the contribution of the other terms will only give subleading corrections.

Let us first discuss the consequences for the CFT dual to pure gravity. In this case the only field is the energy momentum tensor $T$ with $\Delta_\phi=2$. 
By comparing with (\ref{Zathalf}), we immediately find that the partition function converges at least up to $y=1/2$,  since $\sqrt{2} < e^{3.68}$. 
We therefore do not predict a new phase for the third entanglement \ren entropy of pure gravity, which is in agreement with what is generally believed \cite{Headrick:2010zt,Faulkner:2013yia,Hartman:2013mia}. Note however that we cannot fully exclude this because of the order of limit issue that was discussed in section~\ref{ss:NLimit}. We can in fact see that the order of limit plays a role here: From what we have said so far, we would conclude that there is no phase transition at $y=1/2$ either, since the growth of states is slow enough that the partition function converges well beyond that point. Because of crossing symmetry of course we know that this is impossible. The explanation is that the heavy BTZ primaries whose weight grows proportional to $N$ cause this phase transition. Since we took the $N\rightarrow\infty$ limit from the very start, we never took these into account, which is why we do not see the phase transition. In order to see it, we would have to work at finite $N$ \cite{maloneytoappear}.

The situation changes if we add a light scalar to the spectrum. From (\ref{Dfree}) we see that if $\Delta_\phi$ is small enough, the radius of convergence will be smaller than $y=1/2$. That is, there will be a new phase.
The critical value of $\Delta_\phi$ for this to happen is
\be\label{Dcrit}
\Delta_0 \approx \frac{\log 2}{3.68} \approx 0.19\ldots\ .
\ee
Note that we made the simplifying assumption that all the exponents in (\ref{Z2sumdiffexp}) are equal. The more detailed analysis performed in appendix~\ref{app:maxsum} shows that the the result does not change by much under this assumption.

We will discuss the gravity interpretation of these new phase transitions in the discussion section. For the moment, let us simply say a few words about the CFT interpretation.
The results for the generalized free theory may seem problematic. In the context of AdS/CFT, we want to consider the generalized free theory to be some large $N$ limit of a family of honest CFTs. From that point of view, it is of course reasonable that such quantities as the genus 2 partition function can diverge. However, nowhere in the computation of the correlation functions of the generalized free theory did we use the large $N$ limit. We could thus also consider it to be a computation of some standard generalized free theory. If $\Delta_\phi$ is small enough, it may seem worrisome to get a diverging genus 2 partition functions. Note however that generalized free theories do not have a local energy momentum tensor. There is thus no a priori reason why higher genus amplitudes should exist. The one exception is of course the free boson, which is a free theory with a perfectly local energy momentum tensor. In that case we have $h_\phi=1$. This gives in fact exactly the expected radius of convergence: The boundary of the Schottky space $\mathfrak{S}_2$ with $p_1=p_2=x$ gives
\be
Z_2 \sim \sum D(h) x^{h_3}(1-x)^{2h_2}\ ,
\ee
which for $h_i = h/3$ converges worst at $x=1/2$, which is exactly the radius of convergence predicted by the growth $D(h) \sim 2^h$.

\section{The symmetric orbifold point}\label{sec_sym}
\subsection{Growth of OPE coefficients}

Let us now turn to a second class of theories that are interesting for holography: symmetric product orbifold theories. Considering the fact that these are much more complicated than the theories discussed above, it is surprising that in the large $N$ limit we can perform the same computations, and in fact obtain almost identical results. On second thoughts, this is not that surprising after all, since they become free theories in the large $N$ limit, so that similar methods as above can be used. More precisely, states in the symmetric orbifold of some seed theory are given by product of (untwisted) states in the seed theories with products of (twisted) cycles. A ``single trace'' state is then either a single non-vacuum state in a single factor, or a single cycle of length $\geq 2.$ A ``multi trace'' state is a (suitably symmetrized) product of such single trace states. In the large $N$ limit, to order $\cO(1)$  the three point function of three multi trace operators is given by all possible Wick contractions of single trace operators, that is by products of two point functions of the single trace components of the state \cite{Belin:2015hwa}. The correlation functions thus factorize, and the theory is indeed free. To this order in $N$, the computation of the correlation function thus reduces to a counting problem in combinatorics. Let us now work out this counting in detail.

Let us start by considering untwisted sector operators. As described above, the OPE coefficients for these operators will involve OPE coefficients of the seed theory dressed with combinatorics coming from the symmetrization procedure. Let us first introduce the notation.  Denote by $\varphi$ the states in the seed theory, which we choose to be orthonormal. Consider an ordered $K$-tuple $\vec{K}$ of
distinct integers from 1 to $N$, and a $K$-vector $\vec{\varphi}$
of states in the seed theory,
\be
\phi = \phi_{(\vec{K},\vec{\varphi})}\ .
\ee
where the notation is that the factor $K_i$ is in state $\varphi_i$. All other factors are in the vacuum. This is a state of the product theory, which we will refer to as a \textit{prestate}. To obtain a state invariant under the action of $S_N$, we must sum over images of the group. This gives
\be
\Phi = \mathcal{N}_\Phi \sum_{g\in S_N} \phi_{(g.\vec{K},\vec{\varphi})} \, ,
\ee
where $\mathcal{N}_\Phi$ is a normalization constant. These are the untwisted sector states of the symmetric orbifold theory. Computing the OPE coefficients of these states will be in principle difficult. To warm up, we will consider the simplest possible operators, namely those where a single state of the seed theory is picked. Let us call this state $\varphi_1$. Because we are considering the symmetric group, the only other information needed to fully specify a state is the number of times it appears. We will define three operators $\Phi_1,\Phi_2,\Phi_3$ that are made out of $K_1,K_2,K_3$ copies of $\varphi_1$ (without loss of generality, we take $K_1>K_2>K_3$).

The normalization factor was calculated in \cite{Belin:2015hwa} and reads
\be\label{unt_norm}
\mathcal{N}_{\Phi_i} = \big(N! (N-K_i)! K_i !\big)^{-1/2}.
\ee
We now consider the three-point function $\braket{ \Phi_1\,\Phi_2\,\Phi_3 } $. It is easiest to describe the combinatorics using the following diagrammatic representation. The three point function is given by
\begin{eqnarray*}
\Phi_1:&&\overbrace{\underbrace{\bb\bb\bb\bb\bb\bb\bb\bb\bb\bb\bb\bb\bb\bb}_{K_1}\underbrace{\wb\wb\wb\wb}_{K_2-J}\wb\cdots\wb}^N\\
\Phi_2:&&\underbrace{\bb\bb\bb\bb\bb\bb\bb\bb\bb\bb}_J\wb\wb\wb\wb\underbrace{\bb\bb\bb\bb}_{K_2-J}\wb\cdots\wb\\
\Phi_3:&&\underbrace{\bb\bb\bb\bb\bb}_{n_3}\wb\wb\wb\wb\wb\underbrace{\bb\bb\bb\bb\bb\bb\bb\bb}_{K_1+K_2-2J}\wb\cdots\wb
\end{eqnarray*}
Each line in this picture is a prestate, which means that there is also one sum over the group per line to make the operators $S_N$ invariant. It is now easy to keep track of the different combinatorics factors.

Without loss of generality, we can display the states of the first operator, $\Phi_1$, as shown in the picture and the sum over the group gives an overall factor of
\be\label{C_unt_i}
N! \,.
\ee
For the second operator $\Phi_2$, we call $J$ the number of overlapping operators with the first state. There are three contributions that enter. First, we can distribute the $J$ operators among any of the $K_1$ seed operators of $\Phi_1$, minus symmetrisation. Second, we can distribute the $K_2-J$ states over any of the $N-K_1$ vacua, again minus symmetrisation. Finally, we can permute the $N-K_2$ vacua of the second operator and also permute the $K_2$ states in any way. The total contribution from the second sum is then
\be\label{C_unt_ii}
\frac{K_1!}{(K_1-J)!J!} \frac{(N-K_1)!}{(N-K_1-K_2+J)!(K_2-J)!} (N-K_2)!K_2!\,.
\ee
For the third sum in $\Phi_3$, we first call $n_3$ the number of three-point overlaps. These can be distributed over the $J$ overlaps between operators 1 and 2, minus symmetrisation. Then the $K_3$ operators can be permuted in any way among themselves and so can the $N-K_3$ vacua. This gives a total contribution of
\be\label{C_unt_iii}
\frac{J!}{(J-n_3)!n_3!} (N-K_3)!K_3!\,.
\ee
We also have the following relation between $K_i$:
\be\label{C_unt_iv}
K_3=n_3 + K_1+K_2 - 2 J.
\ee
Plugging in equations (\ref{C_unt_i})-(\ref{C_unt_iv}) and the normalization factors (\ref{unt_norm}) in the three-point function, we get the total contribution
\be\label{Nexpansion}
\braket{ \Phi_1 \Phi_2 \Phi_3}=\sum_{n_3=0} \left(c_{111} \right)^{n_3} N^{-n_3/2} C_{123}^{(n_3)},
\ee
where $c_{111}$ is the OPE coefficient of three operators $\varphi_1$ in the seed theory. This formula results from large $N$ factorization. Let us consider the leading term $C_{123}^{(0)}$. It reads
\bea \label{Nzero}
&&C_{123}^{(0)}=\frac{\sqrt{K_1! K_2!K_3!}}{\left(\frac{K_1+K_2-K_3}{2}\right)!\left(\frac{K_1-K_2+K_3}{2}\right)!\left(\frac{-K_1+K_2+K_3}{2}\right)!}\times\\
&&\qquad\qquad\qquad\qquad\qquad\qquad\qquad\times\left( \frac{\sqrt{(N-K_1)!(N-K_2!)(N-K_3)!}}{(N!)^{1/2}(N-1/2(K_1+K_2+K_3))!}\right)\Bigg|_{\mathcal{O}(N^0)}.\nonumber
\eea
Using Stirling's formula (\ref{stirling}), we find that the factor on the second line goes to 1 in the $N\rightarrow\infty$ limit. As promised, we see that we get the exact same expression as for the free theory (\ref{C3ptfree}).

The most general state of the orbifold theory consists of multiple `single trace operators', possibly coming from the twisted sector. We will label such a single trace operator by two indices $i$ and $j$, labeling the length of the cycle and the actual state.  We denote the multiplicity of the single trace state in the full state by $K^{i,j}$. In appendix~\ref{app:symOPE} we show that the three point function of three such states is given by
\be\label{C3ptorb}
C_{123}^{(0)}=
\prod_{i,j}\frac{\sqrt{K_1^{i,j}! K_2^{i,j}!K_3^{i,j}!}}{\left(\frac{K_1^{i,j}+K_2^{i,j}-K_3^{i,j}}{2}\right)!\left(\frac{K_1^{i,j}-K_2^{i,j}+K_3^{i,j}}{2}\right)!\left(\frac{-K_1^{i,j}+K_2^{i,j}+K_3^{i,j}}{2}\right)!}\ .
\ee
For $i$=$j$=1, we have $K^{i,j}_\ell=K_\ell$ and we recover equation (\ref{Nzero}) for the untwisted sector. The general structure is thus simply of a free theory with different types of fields labeled by $i$ and $j$.

\subsection{Contribution to $D(\Delta)$}
Since the three point functions (\ref{C3ptorb}) are the same as in the case of the free field theory (\ref{C3ptfree}), it is clear that the general structure of the genus 2 partition function will look very similar. The key difference is in analyzing the spectrum of the single trace fields. In particular, we want to identify which fields will give the dominant contribution to the growth of $D(\Delta)$ in (\ref{Delta}). There are three possibilities:

\begin{enumerate}
\item
Let $\phi_{\min}$ be the lowest non-vacuum state of the seed theory with weight $\Delta_{\min}$. In the untwisted sector, we can then consider the state with $K$ such seed states, having the total weight $\Delta=K \Delta_{\min}$. This means that we have a contribution of three such states as
\be\label{radconvut}
\sim 2^{\Delta/\Delta_{\min}}\ ,
\ee
the exact same result that we found for the free theory (see equation (\ref{Dfree})). This is the contribution to $D(\Delta)$ of just a single term. On the other hand, we know that the number of untwisted sector states of weight $\Delta$ grows as \cite{Belin:2014fna}
\be
\rho_{ut}(\Delta) \sim \exp \frac{\Delta}{\log \Delta}\ .
\ee
Since this is sub-exponential, the total sum will not give a parametrically larger contribution than (\ref{radconvut}). 

\item
In the twisted sector, the ground state in each cycle of length $n\geq2$ can act as a single trace operator. We can then again choose a state given by a lot of short cycles. The lightest such state is from 2-cycles and has weight
\be\label{Delta_twist}
\Delta_{\text{tw}}= \frac{c}{12}\left(2-\frac12\right)= \frac{c}{8}\ ,
\ee
where $c$ is the central charge of the seed theory. The result is thus the same as in (\ref{radconvut}) with $\Delta_{\min}=\Delta_{\text{tw}}$.

\item
Finally, there are contribution from `long strings', that is states with long cycle lengths. In the cases discussed so far the main contribution came from the three point functions. For long strings, however, the main contribution comes from the fact that there is an exponential number of them, even though their correlation functions are of order 1.
To this end, consider $M$ cycles of length $L$ such that the total length is
\be
L_\text{tot} = M L \,.
\ee
By the argument given in \cite{Belin:2014fna}, this configuration gives a total contribution of $e^{2\pi \Delta}$ states of weight $\Delta$ if $L_{\text{tot}}$ is chosen such that
\be
L_\text{tot}= \frac{12\Delta}{c} \,.
\ee
This then gives
\be
M= \frac{12\Delta}{L c} \,.
\ee
This argument only works if $L\gg 1$. 
Let us now take again a configuration of three such states of $M$ cycles each.
The growth of the OPE coefficients comes from the permutations of the possible $M$ cycles, so for a given triple of states we have the contribution
\be
C_{123}\sim 2^{3M/2}\ 
\ee
to the three-point function. Now we need to take into account the number of such states. For the first state, we get the full Hagedorn number of states $e^{2\pi \Delta_1}$. For the second state we need to pair up half of the cycles with cycles of state 1, which fixes them due to their orthogonality. We can therefore only freely choose the states for the rest of the cycles, giving $e^{\pi \Delta_2}$ possibilities. For the third operator, all states are fixed. Putting everything together and assuming that the maximal contribution comes from $\Delta_1=\Delta_2=\Delta_3=\Delta/3$, we find
\be
2^{3M} e^{\pi \Delta} = e^{\left(\pi + \log 2\frac{ 36}{L c}\right)\Delta}
\ee
Note that because $L\gg 1$, the combinatorics coming from the $M$ different cycles is completely subdominant. So the growth can be well approximated by
\be
e^{\pi \Delta}\ .
\ee

\end{enumerate}

Let us now compare the three different contributions we obtained above. First note that the long string contribution is again universal, just as in the $g=1$ case. That is, it does not depend on the seed theory at all. From the numerical expression we have in \rref{Zathalf}, it appears that long strings cannot be responsible for new phases in the genus two partition function. In fact, in light of how close our numerical estimate is to $\pi$, it is possible that they would exactly saturate the bound for no new phase as in the torus case. The two other contributions however do depend on the seed theory through the weight $\Delta_{\min}$ of the lightest field (\ref{radconvut}) and through its central charge $c$ (\ref{Delta_twist}).

Unlike the genus 1 case, depending on the seed theory, these two non-universal contributions can dominate over the long string. For instance, if we pick $\Delta_1$ small enough, say smaller than $\textstyle\frac{\log 2}{\pi}$, then its contribution will dominate over the long strings. In particular it may lead to a new phase transition before $y=1/2$. Clearly there are seed theories which have such light fields, such as the compactified boson theory with a relatively large (or small) compactification radius. On the other hand one can also find seed theories with no fields light enough, such as the $T^4$ theory near the self dual radii. This shows that the phase diagram of the symmetric orbifold CFTs is not universal.

For concreteness, we can compare these different values for the Ising model (where $c=1/2$ and $\Delta_{\min}=1/8$)
\bea \label{resultsIsing}
y_c^\text{untw}&\sim&0.105 \notag \\
y_c^\text{short}&\sim&0.000411   \,,
\eea
and for the free (compact with self-dual radius) boson  (where $c=1$ and $\Delta_{\min}=1$)
\bea \label{resultsFreeboson}
y_c^\text{untw}&\sim&13.5 \notag \\
y_c^\text{short}&\sim&0.105  \,,
\eea
where in both cases the contributions to the genus 2 partition function (\ref{Z2_nonholom}) are dominated by the twisted sectors.

\section{Conclusion and Discussion}

\subsection{Conclusion}
We have computed the leading contribution to the third mutual \ren information $I^{(3)}$ for the two disjoint intervals in two-dimensional CFTs in the limit of large central charge. In particular, we have investigated phase transitions by studying the radius of convergence in cross-ratio $y$ as we take $c\to\infty$. This was done by mapping the problem to the calculation of a genus two partition function, a quantity sensitive to the full dynamical data of a CFT: both the spectrum and the OPE coefficients. We computed the genus two partition function by performing a Schottky uniformisation of the surface.

Our analysis was simplified as holographic CFTs become generalized free theories in the large central charge limit. This is the case of Einstein gravity (with or without matter) but also of symmetric product orbifolds \cite{Belin:2015hwa}. Large $c$ factorization enabled an exact computation of the OPE coefficients to leading order both for Einstein gravity and for symmetric product theories. We found that in many cases the growth of the OPE coefficients can dominate the partition function. The outcome depends on the conformal dimension of the lightest single-trace operator in the theory. As a consequence, CFTs with such light operators will not exhibit a universal third \ren entropy at large $c$ as new phase transitions can appear before $y=1/2$. The critical bound for the conformal dimension for this to happen was found to be
\be
\Delta\leq \Delta_0 \approx 0.19 \, .
\ee
A CFT with a single-trace operator lighter than that will exhibit a new phase transition at some $y_c<1/2$.

This is the main conclusion of our paper: $I^{(3)}$, and the phase transition it exhibits, are not universal but rather depend on the details of the underlying theory. This is in contrast with $I^{(2)}$ where for symmetric orbifold CFTs the free energy is universal, \emph{i.e.}, only depends on the central charge and is independent of the details of the seed theory, and moreover agrees with the pure gravity free energy. This free energy has a first order phase transition --- at the Hagedorn temperature $T_H=\textstyle\frac1{2\pi}$ --- which is dual to the Hawking-Page transition associated with the dominant saddles \cite{Keller:2011xi}. In anticipation of this it was also conjectured in \cite{Headrick:2010zt} that in the context of the D1-D5 brane system, where the $\mathcal N=(4,4)$ symmetric product orbifold CFT is  believed to correspond to one point on the moduli space of the system, there could potentially exist a non-renormalization theorem for the torus partition function. This suggests that the torus partition function does not acquire corrections at least along some paths on the moduli space of $\mathcal N=(4,4)$ CFTs which connect the orbifold point to the gravity point{\footnote{The conjecture in \cite{Headrick:2010zt} is made for the case of $\mathbb T^4$ compactification.}}.

While our results do not provide further information on this interesting conjecture for genus one, they suggest that for the case of the genus two partition function the observed non-universality rules out the existence of a non-renormalization theorem for large parts of the moduli space: the conformal dimensions of the non-BPS states are not protected by supersymmetry and acquire anomalous dimensions across the moduli space and so the spectrum of the CFT changes from one point to another, which in turn affects the structure of $I^{(3)}$ and its associated phase diagram. Further analysis of the CFTs at other points on the moduli space, even at a perturbative level away from the symmetric product point or other points, might shed further  light on this issue.

Note that even though the \ren entropies disagree, the entanglement entropy may still agree as the new phase that appears at $n=3$ (and presumably at higher $n$) may not leave an imprint as we analytically continue and take $n\to1$. In fact, we know that at $n=2$ the new phase already disappears. Evidence that there are no corrections as $n\to1$ was also found in \cite{Headrick:2010zt} using the replica trick. This is also compatible with the results found in higher dimensions \cite{Belin:2013dva}. In that case, even if there is a new phase above a certain $n_c$, there always exists a neighborhood of $n=1$ where nothing happens, hence the entanglement entropy is not affected by these new phases.

\subsection{Outlook}

\subsubsection*{General genus two Riemann surfaces}

In this paper, we focused on particular Riemann surfaces relevant for computing \ren entropies. In particular the Riemann surface relevant for computing $I^{(3)}$ possesses a $\mathbb Z_3$ symmetry. For this reason, the three complex moduli of the genus two surface only depended on one real parameter: the cross-ratio $y$. A natural question to ask is whether the loss of universality we observed is a general feature of genus two partitions function, with arbitrary moduli. Even though we focused on the  $\mathbb Z_3$ symmetric surfaces, the formula (\ref{Z2quasi}) is valid at arbitrary values of the moduli. It seems very likely that new phase transitions occur at arbitrary values of the moduli. Perturbatively away from the $\mathbb Z_3$ symmetric one, it is certainly true. It would be interesting to check if this continues to be true at arbitrary values of the moduli or not.

\subsubsection*{New gravitational saddles?}
From the bulk point of view our results seem to indicate that a light enough scalar field condenses as we change the moduli of the Riemann surface, thus leading to a new dominant saddle. If we take our bulk theory to be Einstein gravity with a minimally coupled scalar, before the condensation the dominant solution has the scalar set to zero and the manifold is the dominant handle-body described in \cite{Faulkner:2013yia}. Our results suggest that the scalar becomes unstable on the handlebody as we change the moduli, namely that the Laplacian of the scalar acquires a zero-mode at a particular value of the moduli. This might sound surprising at first, since the handlebody solutions are simply quotients of $\mathbb{H}_3$, hence locally AdS. Nevertheless, once a quotient is taken interesting things can happen, as the spectrum of normalizable modes can drastically change. This was for example the case in \cite{Belin:2014lea} for the topological black holes. In that case, taking a quotient of Rindler-AdS produced a manifold that could be unstable to scalar condensation, whereas Rindler-AdS is clearly stable. Our results predict that a similar feature happens for the handlebody solutions. It would be interesting to see if this can be understood from the spectrum of the Laplacian on handlebodies, or whether perhaps the backreacted solutions could be constructed numerically. 

One might also wonder whether these new saddles break replica symmetry in the bulk. It is probably very hard to answer this question without finding the new solutions. In the higher dimensional condensation scenario \cite{Belin:2013dva}, the $\mathbb Z_n$ symmetry was enhanced to a $U(1)$, corresponding to Euclidean (modular) time, and it was never broken in the bulk. It would be interesting to understand whether the situation is analogous here.

\subsubsection*{Higher genus Riemann surfaces}
Another interesting question is the behavior of higher genus partition functions. For simplicity, one can consider only the $\mathbb Z_{n-1}$ symmetric ones relevant for computing $I^{(n)}$. In principle, one can use equation \rref{genusg} to try to estimate the convergence of the partition functions. The difficulty lies in the fact that the relation between the cross-ratio and the Schottky parameters might be complicated. Alternatively, one can consider studying the four point function of twist operators directly along the lines of \cite{Headrick:2010zt}. We argued above that the mechanism driving this phase transition is a scalar condensation. This was the scenario in \ren entropies of spherical regions in higher dimensions \cite{Belin:2013dva}. If indeed the two effects are connected it seems natural to conjecture that the value of $\Delta_0$ will increase as we increase $n$, and asymptote to a given value as $n\to\infty$. It would be very interesting to check this.

\subsubsection*{Order of the phase transition}

Another interesting question it to understand the order of the new phase transitions we find. The Hawking-Page phase transition is always a first order phase transition. If the theory does not have an operator below $\Delta_0$, we then naturally expect a first order phase transition at $y=1/2$. If the theory does have an operator below that bound, there is a new phase transition. 
The intuition from the bulk point of view is that a scalar condenses, which we typically (but not always) associate with a second order phase transition. In the bulk theory, the order of the phase transition is presumably encoded in whether a horizon forms once the scalar backreacts on the geometry. This may happen as soon as the scalar condenses, but it is also possible that the scalar first condenses without creating a horizon. The horizon then forms at some later point as one tunes the moduli. It would be interesting to understand this question better. In the CFT side, this means understanding the phase on the other side of the phase transition which in turn demands understanding precisely how $1/c$ corrections enter. We leave this question for future work.

\subsubsection*{Higher dimensions}

Note that the results we obtain for the growth of the OPE coefficients in generalized free theories apply to any large $N$ CFT in arbitrary dimensions. The only condition we demand is that the theory obeys large $N$ factorization as would be the case for any large $N$ gauge theory or for symmetric product orbifolds in higher dimensions \cite{Belin:2016yll}. It would be interesting to understand whether this exponential growth has any interesting consequences, for example understand whether they play a role in the new phases in the \ren entropies of higher dimensional CFTs \cite{Belin:2013dva}. It is not clear how to compute the replicated manifold partition function in higher dimensions from the dynamical data of the CFT (spectrum and OPE coefficients) but it would be interesting to investigate this.

\section*{Acknowledgments}
We would like to thank Alejandra Castro, Shouvik Datta, Jan de Boer, Ben Freivogel, Matthias Gaberdiel, Sean Hartnoll, Diego Hofman, Nabil Iqbal, Henry Maxfield for useful discussions. We thank Tom Faulkner, Tom Hartman, Matt Headrick and Alex Maloney for helpful remarks on the draft. CAK thanks the Harvard University High Energy Theory Group for hospitality. IGZ thanks Institute of Physics at the University of Amsterdam for hospitality. We thank the Galileo Galilei Institute for Theoretical Physics (GGI) for the hospitality and INFN for partial support during the completion of this work, within the program \textit{New Developments in AdS$_3$/CFT$_2$ Holography}. In addition, IGZ thanks INFN as well as the ACRI (Associazione di Fondazioni e di Casse di Risparmio S.p.a.) for partial support through a YITP fellowship. CAK and IGZ are supported by the Swiss National Science Foundation through the NCCR SwissMAP. AB is supported by the Foundation for Fundamental Research on Matter (FOM). This work is part of the $\Delta$-ITP consortium, a program of the NWO funded by the Dutch Ministry of Education, Culture and Science (OCW).

\appendix 
\section{Appendix}
\subsection{Maximal contribution to $D(\Delta)$}\label{app:maxsum}
For a general $\vec\alpha$, we define
 $H=\alpha_1 \Delta_1+\alpha_2 \Delta_2 +\alpha_3 \Delta_3$, and then write the partition function (\ref{Z2sumdiffexp}) as
\be
Z_2 = \sum_H D_{\vec\alpha}(H) e^{-H}\ ,
\ee
where 
\be
D_{\vec\alpha}(H):= \sum_{H=\alpha_1 h_1+\alpha_2 h_2 +\alpha_3 h_3} |C_{\phi_1\phi_2\phi_3}|^2\ .
\ee 
With $\Delta_i = K_i \Delta_\phi$, we thus want to maximize (\ref{C3ptfree})
subject to the constraint $H=\alpha_1 \Delta_1+\alpha_2 \Delta_2 +\alpha_3 \Delta_3$. After using Stirling's approximation, this can be done using, \emph{e.g.}, Mathematica. In the case of $\alpha_1=\alpha_2=\alpha_3$, one can find an analytic solution, which is indeed $K_1=K_2=K_3$. For general values of $\vec\alpha$ we did this numerically.
In particular we took $\alpha_1=3.68,\alpha_2=3.73$ and $\alpha_3=3.63$. In this case the maximal contribution of $D(H)$ leads to
\be
D(H) \sim e^{0.19\frac{H}{\Delta_\phi}}\ ,
\ee
giving the same critical value as (\ref{Dcrit}).

\subsection{Free fields with derivatives}\label{app:derivatives}
We want to argue here that having derivatives in the free field operators will not increase the growth of their OPE coefficients. For this we consider the following operators
\be
O_k=: \partial^k T \partial^k T:
\ee
We first compute their normalisation. The two-point function gives
\be
\bra{0}\partial^k T \partial^k T:(z)\; :\partial^k T \partial^k T: (0) \ket{0}= \left(\frac{c}{2}\right)^2 2\,\left(\frac{(2k+3)!}{3!}\right)^2 \frac{1}{z^{4k+8}}
\ee
where the $c/2$ comes from the stress tensor contractions, the factor of $2$ from the different possible contractions of $T$'s, and the last factor comes from taking the derivatives. This gives
\be
\mathcal{N}_\mathcal{O} =\left(\frac{c}{2} \sqrt{2}\,\frac{(2k+3)!}{3!}\right)^{-1}
\ee
The OPE coefficient is given by
\be
\mathcal{N}_{\mathcal{O}_1}\mathcal{N}_{\mathcal{O}_2}\mathcal{N}_{\mathcal{O}_3}\bra{0}\partial^{k_1} T \partial^{k_1} T: (0) :\partial^{k_2} T \partial^{k_2} T: (1) :\partial^{k_3} T \partial^{k_3} T: (\infty)\ket{0}
\ee
There is a total factor of $8$ coming from the combinatorics, then there is a factor of $(c/2)^{3}$. The derivatives give a contribution of
\be
\frac{(k_1+k_2+3)!}{3!}\frac{(k_1+k_3+3)!}{3!}\frac{(k_2+k_3+3)!}{3!} \,.
\ee
Taking into account the normalizations, we get a total answer of
\be
\sqrt{8}\;\frac{(k_1+k_2+3)!\,(k_1+k_3+3)!\,(k_2+k_3+3)!}{(2k_1+3)!(2k_2+3)!\,(2k_3+3)!} \,.
\ee
This expression is bounded and should be taken as evidence that adding derivatives does not change the OPE coefficients while it does increase the weight.

\subsection{Untwisted OPE coefficients}\label{app:symOPE}

Let us consider the case of different seed states in the untwisted sector which we will denote by the index $i$, $K_1^i,K_2^i,K_3^i$ being the number of the $i$-th seed theory operator in the total operators. Note that for the contribution to be non-zero, $K_1^i+K_2^i+K_3^i$ must be even and $K_1^i\leq K_2^i+K_3^i$ for all $i$. We are interested in the leading $N$ behavior, so from the start we will only consider configurations that have no triple overlaps, \ie, $n_3=0$. We can think of the different states $i$ as different colors in the diagram below \rref{unt_norm}, so that rather than just black balls we have balls of different colors. The factor from the first state is of course again $N!$. To obtain the second factor, let us first enumerate all possible distributions of the colors. Let $J^i$ be the number of factors of color $i$ that overlap between state 1 and 2. The number of configurations is then
\be
\prod_{i}\binom{K_1^i}{J^i} \frac{(N-K_1)!}{(N-K_1-K_2+J)!}\prod_i\frac{1}{(K_2^i-J_i)!}\ ,
\ee
and for each such configuration we can distribute the states to the colors in
\be
(N-K_2)! \prod_i K_2^i! \ .
\ee
For the last factor, the distribution of colors is completely fixed, since we need to pair up all remaining colors. The only contribution thus comes from the distribution of states to colors, giving $(N-K_3)! \prod_i K_3^i!$. Putting everything together and using $J^i= \frac{1}{2}(K^i_1+K^i_2-K^i_3)$ we get
\bea
&&C_{123}^{(0)}= \left(\frac{\sqrt{(N-K_1)!(N-K_2!)(N-K_3)!}}{(N!)^{1/2}(N-1/2(K_1+K_2+K_3))!}\right)\Bigg|_{\mathcal{O}(N^0)}\times\\
&&\qquad\qquad\qquad\qquad\qquad\qquad\times\;\prod_{i=1}^m\frac{\sqrt{K_1^i! K_2^i!K_3^i!}}{\left(\frac{K_1^i+K_2^i-K_3^i}{2}\right)!\left(\frac{K_1^i-K_2^i+K_3^i}{2}\right)!\left(\frac{-K_1^i+K_2^i+K_3^i}{2}\right)!}.\nonumber
\eea
The first factor goes to 1 in the $N\rightarrow\infty$ limit and so we get
\be
C_{123}^{(0)}=
\prod_{i=1}^m\frac{\sqrt{K_1^i! K_2^i!K_3^i!}}{\left(\frac{K_1^i+K_2^i-K_3^i}{2}\right)!\left(\frac{K_1^i-K_2^i+K_3^i}{2}\right)!\left(\frac{-K_1^i+K_2^i+K_3^i}{2}\right)!}.
\ee
We thus find that the result is indeed the same as that of a free theory with multiple free fields $\Phi_i$.

\subsection{The twisted sector OPE coefficients}

A twisted sector state is slightly more complicated. First, the twisted sector is specified by conjugacy classes of $S_N$, namely a cycle decomposition. Then, each cycle can be in a particular state. The Hilbert space of a cycle is a subset of the original Hilbert space of the seed theory. We will use the following notation
\begin{itemize}
	\item $L^i$ will be the length of the $i^\mathrm{th}$ cycle length
	\item $L$ is the total length of all non-trivial cycles, $L=\sum_{i,j}K^{i,j}L^i$
	\item $n^i$ will be the number of cycles of length $L^i$
	\item $K^{i,j}$ will be the number of cycles of length $L^i$ that are in state $j$. Note that $\sum_j K^{i,j}=n^i$.
	\item $J^{i,j}$ is the number of cycles of length $L^i$ in state $j$ that overlap between state 1 and state 2
	\item $M$ is the total number of factors that overlap between state 1 and state 2, $M= \sum_{i,j}J^{i,j}L^i$.
\end{itemize}
This enables us to calculate the normalization of any twisted sector operator:
\be
\mathcal{N}_\Phi = \sqrt{N!(N-L)!\prod_{i,j} (L^i)^{K^{i,j}} K^{i,j}!} \, .
\ee
Note that for this to include the most general operator, it can include untwisted sector factors, namely those where $L^i=1$ but $j\neq0$.

We now describe the different combinatorial factors that can appear in the three-point function. We will only consider the order $N^0$ contribution to the OPE coefficient as the corrections depend on multipoint function of twist operators which are harder to compute. Nevertheless, we have already shown in \cite{Belin:2015hwa} that the form \rref{Nexpansion} remains valid, although the coefficients are no longer simply related to the OPE coefficients of the seed theory.

The first sum gives as usual a factor of
\be
N!
\ee
For the second sum we will again first distribute the `colors', and then distribute the states to the colors. In this we can mimic the untwisted case. The first step gives
\be
\prod_{i,j}\binom{K_1^{i,j}}{J^{i,j}}\,\frac{(N-L_1)!}{(N-L_1-L_2+M)!}\prod_i\frac{1}{(K_2^{i,j}-J_{i,j})!}\ .
\ee
The first factor comes from matching up the $J^{i,j}$ overlapping colors of state 2 with the $K_1^{i,j}$ corresponding colors of state 1. The second factor comes from distributing all remaining factors, and the last factor eliminates the overcounting from the identical colors $(i,j)$. The second step gives
\be\label{fac2}
(N-L_2)! \prod_{i,j} K_2^{i,j}!\,(L^i)^{J^{i,j}} \ .
\ee
The extra factor $(L^i)^{J^{i,j}}$ comes from the fact that we can cyclically shift each cycle of state 2 that matches a cycle of state 1. Note that we do not have such a contribution for the non-matching cycles of state 2, since those we already fully permuted in the factor $\frac{(N-L_1)!}{(N-L_1-L_2+M)!}$.
Finally the third sum gives the same factor as (\ref{fac2}),
\be
(N-L_3)! \prod_{i,j} K_3^{i,j}!\,(L^i)^{K_3^{i,j}} \ .
\ee
We have no choice of lining up the colors, since we need to match all unmatched ones. The distribution of states to colors then gives the same answer as before, with the cyclic permutations taken into account.
Gathering all the factors and using $M= \frac{1}{2}(L_1+L_2-L_3)$ and
$J=\frac{1}{2}(K_1+K_2-K_3)$ we can write the final result in symmetric form,
\begin{multline}
\left(\frac{\sqrt{(N-L_1)!(N-L_2!)(N-L_3)!}}{(N!)^{1/2}(N-1/2(L_1+L_2+L_3))!}\right)\Bigg|_{\mathcal{O}(N^0)}\times \\
\times\;\prod_{i,j}\frac{\sqrt{K_1^{i,j}! K_2^{i,j}!K_3^{i,j}!}}{\left(\frac{K_1^{i,j}+K_2^{i,j}-K_3^{i,j}}{2}\right)!\left(\frac{K_1^{i,j}-K_2^{i,j}+K_3^{i,j}}{2}\right)!\left(\frac{-K_1^{i,j}+K_2^{i,j}+K_3^{i,j}}{2}\right)!}\
\end{multline}
Note that in particular the $L^i$ drop out.
The result is thus simply a generalization of the untwisted sector result, namely
\be
C_{123}^{(0)}=
\prod_{i,j}\frac{\sqrt{K_1^{i,j}! K_2^{i,j}!K_3^{i,j}!}}{\left(\frac{K_1^{i,j}+K_2^{i,j}-K_3^{i,j}}{2}\right)!\left(\frac{K_1^{i,j}-K_2^{i,j}+K_3^{i,j}}{2}\right)!\left(\frac{-K_1^{i,j}+K_2^{i,j}+K_3^{i,j}}{2}\right)!}\ .
\ee

\bibliographystyle{ytphys}
\bibliography{ref}

\end{document}